\documentclass[%
    aps,
    prd,
    letterpaper,
    superscriptaddress,
    preprintnumbers,
    floatfix,
    twocolumn,
    nofootinbib,
    showpacs,
    hyphens
]{revtex4-1}  

\usepackage{subfigure}
\usepackage{graphicx}% Include figure files
\usepackage{dcolumn}% Align table columns on decimal point
\usepackage{bm}% bold math
\usepackage{xcolor}
\usepackage{amsmath,amsfonts,amssymb,amsthm}
\newcommand{\ket}[1]{\left| #1 \right>} % for Dirac bras
\usepackage{slashed}
\usepackage{hyperref}% add hypertext capabilities

\usepackage{multirow}

\newcommand{\bs}{\boldsymbol}

\newcommand{\Ham}{H}
\newcommand{\Vop}{V}
\newcommand{\Top}{T}

\begin{document}

\preprint{FERMILAB-PUB-26-0061-T}

\title{Renormalon subtracted nonrelativistic QCD for heavy hadron systems}
\author{Beno\^{i}t Assi}
\affiliation{Department of Physics, University of Cincinnati, Cincinnati, Ohio 45221, USA}
\affiliation{Theory Division, Fermi National Accelerator Laboratory, Batavia, IL, 60510}
\author{Andreas S. Kronfeld}
\affiliation{Theory Division, Fermi National Accelerator Laboratory, Batavia, IL, 60510}
\author{Simon Vaiva}
\affiliation{Theory Division, Fermi National Accelerator Laboratory, Batavia, IL, 60510}
\affiliation{Aix Marseille Univ, Université de Toulon, CNRS, CPT, Marseille, France}
\author{Michael~L.~Wagman}
\affiliation{Theory Division, Fermi National Accelerator Laboratory, Batavia, IL, 60510}

\date{\today}

\begin{abstract}
We present a renormalon-subtracted formulation of potential nonrelativistic QCD (pNRQCD) for precision spectroscopy of heavy hadron systems, combining variational and Green's function Monte Carlo (VMC/GFMC) methods with static two- and three-body potentials through NNLO in $\alpha_s$ (leading order in $1/m_Q$). Minimal renormalon subtraction (MRS) systematically sums leading factorially growing terms, thereby stabilizing perturbative convergence and reducing renormalization-scale dependence. We tune charm and bottom quark masses to spin-averaged $1S$ quarkonium states and predict $\Omega_{ccc}$, $\Omega_{ccb}$, $\Omega_{cbb}$, and $\Omega_{bbb}$ baryon masses, as well as QCD-stable baryons containing top quarks. NNLO MRS results undershoot lattice QCD by 125--175~MeV, a difference we attribute to contributions beyond the static perturbative potentials, including $1/m_Q$ corrections. Applying these methods to unequal-mass fully-heavy tetraquarks, we determine the critical heavy-to-light mass ratio for binding and compute binding energies across the mass-ratio landscape.
\end{abstract}

\maketitle

\section{Introduction}

Heavy-quark bound states provide theoretically controlled laboratories for testing QCD dynamics across multiple scales~\cite{Gross:1973id,Politzer:1973fx,Gross:1974jv,Jones:1974mm,Caswell:1974gg,Tarasov:1980au,Larin:1993tp,vanRitbergen:1997va,Zoller:2016sgq,Baikov:2016tgj,Herzog:2017ohr,Luthe:2017ttc}.
The hierarchy of scales $m_Q \gg m_Q v \gg m_Q v^2$ characteristic of nonrelativistic heavy quarks enables systematic effective field theory (EFT) treatments~\cite{Appelquist:1977es,Caswell:1985ui,Lepage:1992tx, Brambilla:1999qa}, with potential nonrelativistic QCD (pNRQCD)~\cite{Pineda:1997bj,Pineda:2000gza,Brambilla:2004jw,Brambilla:2006wp,Brambilla:2009bi,Pineda:2011dg} organizing the dynamics through a Schr{\"o}dinger equation with potentials matched order-by-order in $\alpha_s$ and $1/m_Q$.
For systems containing multiple heavy quarks -- quarkonia, triply-heavy baryons, and fully-heavy tetraquarks -- pNRQCD predictions should  provide a controlled expansion for low-energy observables when $m_Q \gg \Lambda_{\rm QCD}$, as the heavy quark mass suppresses relativistic corrections and the velocity expansion converges systematically~\cite{Caswell:1985ui,Lepage:1992tx}.

In our previous works~\cite{Assi:2023cfo,Assi:2023dlu,Assi:2025ysr}, two of us employed pNRQCD and quantum Monte Carlo (QMC) techniques to study heavy-hadron spectra at fixed order in $\alpha_s$.
Those calculations revealed a significant obstacle to precision predictions: renormalization-scale variation induced shifts of order $100$~MeV in binding energies -- comparable to the target precision needed to match lattice-QCD benchmarks~\cite{Meinel:2010pw,Brown:2014ena,Mathur:2018epb,Bazavov:2019qoo, Brambilla:2021wqs,Mathur:2022nez,Brambilla:2022het} -- and successive perturbative orders showed no sign of convergence. 
This behavior is characteristic of renormalon ambiguities~\cite{Beneke:1994sw,Bigi:1994em,Beneke:1994rs,Luke:1994xd,Kronfeld:1998di,Lee:1996yk, Lee:1999ws,Komijani:2017vep,Sumino:2020mxk}, which arise from factorial growth of perturbative coefficients~\cite{Bender:1971gu,Bender:1973rz, Lautrup:1977hs,tHooft:1977xjm,Brown:1992pk} and manifest as enhanced $\mu$~dependence~\cite{Beneke:1998ui}. 
For multiquark systems, where delicate cancellations among color channels amplify sensitivity to the static potential~\cite{Billoire:1979ih,Fischler:1977yf,Peter:1997me,Schroder:1998vy,Anzai:2009tm,Smirnov:2008pn,Smirnov:2009fh}, the problem is particularly acute.

This paper addresses the renormalon problem by implementing minimal renormalon subtraction (MRS)~\cite{Komijani:2017vep,Brambilla:2017hcq,Kronfeld:2023jab,Kronfeld:2024qao,Kronfeld:2025ayg} 
within our pNRQCD-QMC framework. MRS reorganizes the perturbative series by isolating 
the renormalon into an explicit power 
correction~\cite{Falk:1992fm,Falk:1992wt,Bigi:1993zi,Beneke:1994sw,Bigi:1994em,Beneke:1994rs,Luke:1994xd} and resumming the leading factorial growth, 
leading to substantially improved convergence 
and reduced scale dependence. The result is a dramatic stabilization: scale variation in meson and baryon binding energies
decreases by factors of 3--10, order-by-order shifts shrink by similar factors, 
and uncertainty bands overlap across leading order (LO), next-to-leading order 
(NLO), and next-to-next-to-leading order (NNLO) -- hallmarks of a well-behaved 
perturbative treatment. While some renormalon-subtracted schemes have been 
successfully applied to heavy quarkonium~\cite{Hoang:2000fm,Pineda:2001zq,
Ayala:2014yxa,Hoang:2008yj,Tormo:2013tha,Ayala:2020odx,Takaura:2018vcy,
Bazavov:2019qoo},
MRS has only been used to determine quark masses for all flavors except top~\cite{Brambilla:2017hcq,FermilabLattice:2018est}, and work is underway to use the MRS static potential to determine $\alpha_s$ from lattice-QCD data~\cite{Brambilla:2022het, Leino:2025pvl}.
This work is the first application of MRS to bound-state problems in pNRQCD.

Our implementation applies the MRS procedure to the LO/NLO/NNLO static two-body 
potentials~\cite{Fischler:1977yf,Peter:1997me,Schroder:1998vy,Anzai:2009tm,Smirnov:2008pn,Smirnov:2009fh,Pineda:2000gza,Brambilla:2006wp,Brambilla:2009bi,Brambilla:1999qa,Sumino:2020mxk} and include fixed-order static three-body potentials arising at NNLO~\cite{Brambilla:2009cd,Assi:2023cfo}.
We choose the renormalization scale to interpolate smoothly between the natural scale and a fixed scale (Sec.~\ref{app:mrs_impl}) 
to avoid derivative discontinuities and to ensure stable stochastic sampling. 
Ground-state energies are extracted via variational Monte Carlo (VMC) and Green's function Monte Carlo (GFMC) projections from Coulombic trial wavefunctions. %systematically improvable

We validate the framework by tuning charm and bottom quark masses to reproduce spin-averaged $1S$ quarkonium binding energies~\cite{Hoang:2000fm,Pineda:2001zq,Ayala:2014yxa,Hoang:2008yj,FermilabLattice:2018est,Tormo:2013tha,Ayala:2020odx}, then predict masses for triply-heavy baryons $\Omega_{ccc}$, $\Omega_{ccb}$, $\Omega_{cbb}$, and $\Omega_{bbb}$ that can be directly compared with lattice QCD.
At NNLO with MRS, our predictions undershoot lattice-QCD results~\cite{Brown:2014ena,Meinel:2010pw,Mathur:2018epb,Mathur:2022nez,Bazavov:2019qoo,Brambilla:2021wqs,Brambilla:2022het} by 125--175~MeV across all channels -- approximately 1--4\% of the baryon masses -- with fractional discrepancies scaling as $\sim 1/m_Q$. This scaling is consistent with the expected size of neglected $\mathcal{O}(1/m_Q)$ velocity-dependent corrections in the static potential and $\mathcal{O}(1/m_Q^2)$ spin-orbit, spin-spin, and Darwin corrections included in lattice-NRQCD actions but absent from our static potentials. Critically, MRS reduces FO scale-variation bands by roughly two orders of magnitude (e.g., from $\sim 700$~MeV to $\sim 10$~MeV at NNLO for $\Omega_{bbb}$) and reduces LO-to-NLO shifts by a factor of $\sim 5$--10 in baryon masses, signaling effective removal of the leading renormalon contribution.

While the top quark decays before forming long-lived hadrons, our framework nonetheless 
provides well-defined predictions for top-containing systems that serve multiple purposes. 
First, the $t\bar{t}$ binding energy, mixed-flavor $t\bar{c}$ and $t\bar{b}$ mesons, and 
top-containing baryons such as $\Omega_{tcc}$, $\Omega_{tbb}$, and $\Omega_{ttt}$ offer 
benchmarks for testing heavy-quark dynamics in the $m_Q \to \infty$ limit. 
Second, near-threshold Coulombic effects in $t\bar{t}$ production modify 
differential cross sections and spin correlations at hadron 
colliders~\cite{Fuks:2024yjj,Fuks:2021xje,Garzelli:2025toponiumLHC,maltoni2024quantumdetectionnewphysics} 
-- indeed, recent LHC data show an excess of more than $5\sigma$ at the $t\bar{t}$ threshold, 
consistent with toponium formation~\cite{CMS:2025kzt,ATLAS:2026ttThreshold}, with angular correlations 
favoring pseudoscalar quantum numbers.  
Third, the top-containing tetraquarks $bb\bar{t}\bar{t}$, $bc\bar{t}\bar{t}$, and 
$cc\bar{t}\bar{t}$ complete the mapping of the all-heavy tetraquark binding landscape 
across quark flavors.

Extending to fully-heavy tetraquark states $(QQ\bar{Q}'\bar{Q}')$, we map the binding landscape as a function of heavy-to-light mass ratio $m_Q/m_{Q'}$ and determine the critical ratio above which deeply bound configurations emerge. For physical quark masses, we find bound $bb\bar{t}\bar{t}$, $bc\bar{t}\bar{t}$, and $cc\bar{t}\bar{t}$ states. No binding is observed for equal-mass or near-equal-mass combinations such as $bb\bar{b}\bar{b}$ or $cc\bar{b}\bar{b}$. 

Further extensions of the MRS-pNRQCD framework to excited states and QCD-like dark sectors are straightforward, as shown for the fixed-order case in Ref.~\cite{Assi:2023cfo}.
For composite dark matter models with heavy dark quarks such as those in Refs.~\cite{Asadi:2021yml,Asadi:2021pwo,Kribs:2009fy,Kribs:2016cew}, this approach enables efficient parameter-space scans of dark hadron spectra without requiring dedicated lattice-QCD simulations at each mass point. 

MRS is one member of a broader class of renormalon-subtracted short-distance schemes; widely used alternatives include the potential-subtracted (PS), 1S, MSR, and renormalon-subtracted (RS/RS$'$) schemes~\cite{Beneke:1998ui,Hoang:1999sp,Hoang:2008yj,Pineda:2001zq}. A practical advantage of MRS is that it makes the leading renormalon contribution fully explicit -- by isolating and summing the factorial growth into a well-defined term. The remaining short-distance coefficients exhibit transparent power counting and markedly improved $\mu$ stability while maintaining a direct connection to standard short-distance quark-mass inputs~\cite{Komijani:2017vep,Brambilla:2017hcq,Kronfeld:2023jab,Kronfeld:2024qao,Kronfeld:2025ayg}.

The remainder of this paper is organized as follows. Section~\ref{sec:qcd_mesons} presents quarkonium results and quark mass determinations.
Section~\ref{sec:methods} gives a quick overview of pNRQCD and MRS.
Section~\ref{sec:qcd_baryons} extends the analysis to triply-heavy baryons and compares with lattice QCD.
Section~\ref{sec:qcd_tetras} explores fully-heavy tetraquark binding and critical mass ratios.
Section~\ref{sec:out} discusses prospects for including $1/m_Q$ corrections, spin-dependent dynamics, and applications beyond QCD.
Technical details of the MRS implementation and QMC methodology are provided in an Appendix.

\section{Framework and implementation}
\label{sec:methods}

This section outlines the pNRQCD formalism, quantum Monte Carlo methods~\cite{Assi:2023cfo}, and MRS implementation~\cite{Kronfeld:2023jab} used throughout this work.
Full technical details, including explicit potential forms and MRS derivations, are provided in the Appendix and in Refs.~\cite{Assi:2023dlu,Assi:2025ysr,Brambilla:2017hcq,Kronfeld:2024qao,Kronfeld:2025ayg}.

\subsection{Potential nonrelativistic QCD}
\label{sec:pnrqcd}

The pNRQCD Hamiltonian for systems of heavy quarks and antiquarks is organized as an expansion in $1/m_Q$ and $\alpha_s$, with dynamics encoded in a Schr{\"o}dinger equation containing matched static potentials~\cite{Brambilla:2004jw,Pineda:2011dg,Brambilla:1999xf}. At leading order in $1/m_Q$, the Hamiltonian takes the form
\begin{equation}
  \Ham = \Top + \Vop^{\psi\chi} + \Vop^{\psi\psi} + \Vop^{\chi\chi} + \mathcal{O}(1/m_Q),
\end{equation}
where $\Top$ is the nonrelativistic kinetic energy operator for heavy quark fields $\psi(\bs{r})$ and antiquark fields $\chi(\bs{r})$, and the potential operators encode color-dependent interactions between quarks, antiquarks, and quark-antiquark pairs. The two-body potentials $V^{\psi\chi}$, $V^{\psi\psi}$, and $V^{\chi\chi}$ are known through NNLO in $\alpha_s$~\cite{Kniehl:2002br,Anzai:2013tja,Brambilla:2009cd,Assi:2023cfo}, while three-body and four-body potentials enter at NNLO and are detailed in Refs.\cite{Brambilla:2009cd,Assi:2023cfo,Assi:2025ysr}

For quarkonium states, only the color-singlet quark-antiquark potential contributes:
\begin{equation}\label{eq:Vsing}
    V^{\psi\chi}_{\mathbf{1}}(r,\mu) = -\frac{C_F \alpha_V(r,\mu)}{r} ,
\end{equation}
where $C_F = 4/3$ and the effective coupling $\alpha_V(r,\mu)$ encodes higher-order corrections to the Coulomb form~\cite{Fischler:1977yf,Peter:1997me, Schroder:1998vy, Anzai:2009tm, Smirnov:2008pn, Smirnov:2009fh, Pineda:2000gza}. At LO, $\alpha_V(r,\mu) = \alpha_s(\mu)$ exactly. For triply-heavy baryons, the color-antisymmetric quark-quark potential dominates, while tetraquark configurations involve intricate cancellations among multiple color channels, as detailed in Refs.~\cite{Huang_2021,Assi:2023dlu}.

Spin-dependent operators, suppressed in the nonrelativistic expansion by powers of $1/m_Q$, are not included in the present work~\cite{Brambilla:2004jw,Pineda:2011dg,Brambilla:2009bi}. 
This approximation is well controlled for bottom quarks but of limited accuracy for charm. In standard nonrelativistic power counting, spin-dependent (SD) effects generate energy shifts of order $\Delta E_\text{SD}\sim m_Q v^4$, i.e., they are suppressed by $v^2$ relative to the leading binding energy $E_\text{bind}\sim m_Q v^2$~\cite{Brambilla:2004jw,Pineda:2000gza}.
For Coulombic systems, where $v\sim\alpha_s$~\cite{Caswell:1985ui,Lepage:1992tx}, this implies $\Delta E_\text{SD}/E_\text{bind}\sim\alpha_s^2$~\cite{Brambilla:2004jw}, which is $\sim$5\% for bottom and $\sim$10\% for charm. 
Nevertheless, the size of these corrections should be validated through comparison with lattice-QCD calculations that include spin-dependent and $1/m_Q$ effects. Throughout, LO, NLO, and NNLO refer to the order in $\alpha_s$ of the static potentials only. In the standard simultaneous expansion of pNRQCD in $\alpha_s$ and $1/m_Q$, mass corrections already enter at $\mathcal{O}(\alpha_s^2)$. These, together with the $1/m_Q^2$ potentials, are not included here.

\subsection{Implementation of minimal renormalon subtraction}
\label{app:mrs_impl}

Fixed-order perturbative calculations of the static potential exhibit factorial growth of coefficients at high orders, generating renormalon ambiguities that manifest as severe renormalization-scale dependence~\cite{Beneke:1998ui}. At NNLO, varying $\mu$ by a factor of two induces $\sim 800\,\text{MeV}$ shifts in binding energies for triply-heavy baryons.

The MRS framework~\cite{Brambilla:2017hcq,Kronfeld:2023jab} reorganizes the perturbative series by isolating the
leading % $u$ is never explained in this paper, and we don't need to
renormalon into an explicit nonperturbative power correction $\Lambda/Q$, while the remaining short-distance coefficients exhibit substantially reduced $\mu$ dependence. The key steps are:

\begin{enumerate}
\item \textit{Series reorganization}:
For a generic potential in representation $R$, pNRQCD tells us that $\mathcal{V}_R(r)=-(C_R/r)\sum_{n=0} v_n(r,\mu) \alpha_s(\mu)^{n+1} +\Lambda_R$, with a perturbative series and a constant of order $\Lambda_\text{QCD}$. Taking a derivative with respect to $r$ removes the constant,
yielding a renormalon-subtracted series with coefficients $f_n(\mu)$ related to the original $v_n(\mu)$ via
\begin{equation}\label{eq:fl}
    f_n = v_n - \frac{2}{p} \sum_{k=0}^{n-1} (k+1) \beta_{n-1-k} v_k,
\end{equation}
where $\beta_k$ are the $\beta$-function coefficients and $p=1$ for the binding energy.

\item \textit{Borel resummation}: The tail of the series ($n \geq L$, where $L$ is the number of available coefficients) is summed using a Borel integral, yielding a convergent representation
\begin{equation}
    V_\text{B}(r,\mu) = -\frac{C_R}{r} \frac{V_0}{2\beta_0} \mathcal{J}(b, 1/2\beta_0\alpha_s(\mu)),
\end{equation}
where $V_0$ is obtained via a convergent sum containing the $f_n$, $\mathcal{J}(c,y) = e^{-y}\Gamma(-c)\gamma^\star(-c,-y)$ and $\gamma^\star$ is the limiting function of the incomplete gamma function (see the Appendix for details).

\item \textit{Power correction}: The constant $\Lambda_R$ is now free of the leading renormalon ambiguity. In this work, it is in practice absorbed into the quark masses for the singlet potential and neglected in other cases.
\end{enumerate}

For the static potential, we implement MRS at each evaluation point $r$ by constructing the renormalon-subtracted perturbative series through NNLO. 
This is achieved by subtracting the asymptotic renormalon contributions $V_n$ from the fixed-order coefficients $v_n$ and supplementing the result with the Borel-resummed tail, together with the corresponding explicit power correction. 
The renormalization scale is chosen dynamically as $\mu(r)=1/r$ in the perturbative regime and smoothly transitioned to a fixed $\mu_\text{cut}$ for $r>1/\mu_\text{cut}$ using the prescription
\begin{equation}
    \mu'(r;\lambda) = \sqrt{\mu_{\text{cut}}^2 + \lambda^2/r^2},
    \label{eq:mu_smooth}
\end{equation}
where $\lambda$ is a dimensionless scale parameter. The value $\lambda=1$ is used for all central results; the residual MRS scale dependence is probed by varying $\lambda \in \{1/2, 1, 2\}$. This prescription avoids discontinuities in the derivative of the potential, ensures stable QMC sampling, and exhibits improved convergence relative to hard-cutoff schemes.
As discussed in the Appendix, choosing $\mu_\text{cut}=1$~GeV ensures that MRS is deployed, in practice, in the bound-state problem.

The MRS static potential at order $L$ in the coupling is
\begin{align}\label{eq:V_MRS}
    V_\text{MRS}(r) = -\frac{C_R}{r}\bigg[
      &\sum_{l=0}^{L-1}\bigl(v_l - V_l\bigr)\,\alpha_s^{l+1}\bigl(\mu'(r)\bigr)
      \notag\\
      &+ \frac{V_0}{2\beta_0} \mathcal{J}\left(b,\,1/2\beta_0\alpha_s(\mu'(r))\right)
    \bigg],
\end{align}
where $C_R$ is the color factor for the relevant representation ($C_F = 4/3$ for singlet $Q\bar{Q}$, $C_B = C_F/2 = 2/3$ for the color-singlet $QQ$ pair in a baryon, etc.), $v_l$ are the fixed-order potential coefficients evaluated at scale $\mu'(r)$, $V_l = V_0(2\beta_0)^l\Gamma(b{+}l+1)/\Gamma(b+1)$ are the asymptotic (renormalon) contributions with $b = \beta_1/(2\beta_0^2)$, and $V_0$ is determined from the renormalon-subtracted coefficients $f_l$ via Eq.~\eqref{eq:fl}. The truncated sum in Eq.~(\ref{eq:V_MRS}) is the difference between the truncated perturbative series and the truncated asymptotic series; the second is the Borel-resummed tail. Together they provide a scheme in which renormalon cancellation is exact order by order, leaving residual scale dependence that is genuinely perturbative.

We evaluate $V_0$ from the renormalon-subtracted coefficients $f_l$ via Eq.~(\ref{eq:V0}) truncated at NNLO ($L=3$): $V_0 \approx 0.771$ for the singlet potential ($n_f=4$, $\overline{\rm MS}$), with $\lesssim 0.1\%$ relative shift to $\approx 0.772$ at N3LO, indicating well-controlled perturbative convergence.
Throughout, the number of active flavors is set by the lightest valence quark of each state -- $n_f=3$ for charm-containing systems, $n_f=4$ for bottom systems without charm, and $n_f=5$ for top -- so that the coupling, the potential coefficients, and the renormalon subtraction are evaluated consistently (the $n_f=4$ above corresponds to the bottom systems).
The same assignment follows directly from the reduced mass, the only mass scale that enters the dynamical calculation and hence sets the scale at which the coupling and subtraction are evaluated. For all states considered here it falls into the windows fixed by the heavy-quark thresholds: below the bottom-quark mass gives $n_f=3$, between the bottom- and top-quark masses $n_f=4$, and equal to the top-quark mass $n_f=5$---an equivalent restatement of the criterion above.
In the fixed-order scheme the active-flavor number instead follows the approach detailed in Ref.~\cite{Assi:2023cfo}.
The difference $V_0^\text{N3LO}-V_0^\text{NNLO}\approx 0.001$ is itself an N3LO effect and is subsumed into the overall N3LO truncation uncertainty rather than tracked as a separate source of error.

For multiquark systems such as baryons and tetraquarks, the same MRS construction applies to each two-body color channel, with only the overall color factor $C_R$ changed; the spatial structure and Borel resummation are identical. At NNLO, color-dependent contributions from non-Abelian (H-topology) diagrams shift the singlet $v_2$ coefficient by representation-dependent amounts $\delta_R \propto C_R(\pi^4 - 12\pi^2)$~\cite{Assi:2023cfo}. We apply the same MRS Borel resummation Eq.~\eqref{eq:V_MRS} to all color channels, using the shifted $v_2$ coefficients.
This is strictly justified for the singlet, where the pole-mass renormalon cancellation is established~\cite{Kronfeld:2023jab, Kronfeld:2024qao}. For non-singlet channels (octet, antisymmetric, symmetric), applying MRS identically is a scheme choice; a careful derivation of how the renormalon structure extends to these representations is left to future work.
Genuine three- and four-body potentials also arise at NNLO; these are included at fixed order (using the same scheme and scale for $\alpha_s$ as in the two-body potential), as the MRS resummation formalism for many-body potentials is not yet developed.

To probe perturbative scale dependence in MRS, we rescale $\mu'$ via $r \to r/\lambda$ with $\lambda \in \{1/2, 1, 2\}$, equivalent to varying the effective renormalization scale by a factor of two around its central value. For fixed-order calculations, the conventional scale variation $\mu \in [\mu_p/2, 2\mu_p]$ with $\mu_p = 4\alpha_s(\mu_p) m_Q$ is used. The resulting scale band in either scheme is not a quantitative estimate of perturbative truncation error; it captures residual scale dependence as a diagnostic of scheme stability, and order-by-order shifts are a more direct gauge of the remaining truncation.

The full MRS formalism, including derivations of the reorganized coefficients and Borel integrals, is presented in the Appendix.

\subsection{Quantum Monte Carlo methods}
\label{sec:qmc}

We compute binding energies using variational Monte Carlo (VMC) and Green's function Monte Carlo (GFMC) as detailed in Refs.~\cite{Carlson:2014vla,Gandolfi:2020pbj,Assi:2023cfo}. VMC minimizes the energy expectation value
\begin{equation}
  \Delta E_{\text{VMC}} = \min_{\bs{\omega}} \frac{\langle \Psi_T(\bs{\omega}) | \Ham | \Psi_T(\bs{\omega}) \rangle}{\langle \Psi_T(\bs{\omega}) | \Psi_T(\bs{\omega}) \rangle}
\end{equation}
over a family of trial wavefunctions $\Psi_T(\bs{R}; \bs{\omega})$ with parameters $\bs{\omega}$. The matrix elements are evaluated stochastically by sampling configurations $\bs{R}$ from the probability distribution $|\Psi_T(\bs{R})|^2$ using the Metropolis algorithm.

GFMC projects out the exact ground state by evolving the trial wavefunction in imaginary time $\tau$:
\begin{equation}
    \lim_{\tau \to \infty} e^{-\Ham \tau} | \Psi_T \rangle \propto | \Psi_0 \rangle,
\end{equation}
where $|\Psi_0\rangle$ denotes the true ground state. 
In this asymptotic limit, excited-state contributions are exponentially suppressed, and the projection becomes independent of the detailed structure of the trial wavefunction provided it has nonvanishing overlap with $|\Psi_0\rangle$.

The imaginary-time propagator is implemented using the Trotter-Suzuki decomposition~\cite{Carlson:2014vla}:
\begin{equation}
    e^{-\Ham \delta\tau} \approx e^{-V \delta\tau/2} e^{-\Top \delta\tau} e^{-V \delta\tau/2} + \mathcal{O}(\delta\tau^3),
\end{equation}
with the kinetic term generating Gaussian diffusion and the potential term entering as a multiplicative weight.
Here, $\delta\tau = \tau / N_{\rm steps}$ relates the Trotterization scale to the number of steps $N_{\rm steps}$ for which GFMC evolution is performed.

For finite projection time $\tau$, the effective energy estimator
\begin{equation}
  \Delta E(\tau) = \frac{\langle \Psi_T | \Ham e^{-\Ham \tau} | \Psi_T \rangle}
  {\langle \Psi_T | e^{-\Ham \tau} | \Psi_T \rangle}
\end{equation}
approaches the ground-state energy monotonically from above. 
This property follows from the spectral decomposition of the propagator and ensures that $\Delta E(\tau)$ provides a strict variational upper bound at any finite $\tau$, with convergence controlled by the energy gap to the lowest excited states.

Statistical uncertainties are estimated via bootstrap resampling. 
Systematic effects associated with the imaginary-time projection are quantified through variations of the projection time $\tau$ as detailed in Ref.~\cite{Assi:2023cfo}. 
Residual biases arise from finite-$\tau$ extrapolation and discretization errors in $\delta\tau$, which are assessed by repeating the calculations over a range of projection times and time steps, following standard GFMC practice~\cite{Kalos:1962,Ceperley:1980,Carlson:2015}. For quarkonium, GFMC results were compared with conventional quantum-mechanical perturbation theory in Ref.~\cite{Assi:2023cfo}. 

\section{Doubly-heavy mesons}\label{sec:qcd_mesons}

\label{sec:QQbar_trial}

The static potential receives logarithmic perturbative corrections which are conveniently absorbed into the scheme-dependent coupling $\alpha_V(r,\mu)$ defined by Eq.~\eqref{eq:Vsing}. 
In this regime the exact ground-state wavefunction is close to the Coulombic form over the physically relevant distance range. 
We therefore adopt a Coulombic ansatz,
\begin{equation}
    \Psi_T^{Q\bar Q}(\mathbf r_{12}) = \frac{1}{\sqrt{\pi}a_0^{3/2}} e^{-|r_{12}|/a_0},
\end{equation}
where the Bohr radius is set by evaluating the MRS-resummed potential at a reference separation $r^*$:
\begin{equation}
  a_0 = \frac{2}{C_R\, \alpha_V^{\rm MRS}(r^*)\, m_Q},
  \label{eq:aL_trial}
\end{equation}
with $r^* = 0.35\,\text{GeV}^{-1}$, a matching scale chosen in the perturbative regime ($1/r^* \approx 2.9$~GeV) where the MRS resummation is meaningfully active above the IR cutoff $\mu_\text{cut} = 1$~GeV. Since GFMC projection from any reasonable trial wavefunction converges to the same ground state, this choice affects only the variational efficiency, not the extracted binding energies.

\subsection{Quark mass extraction and binding energy scaling}

At the order we work, the Hamiltonian is
\begin{equation}
    H = -\sum_{i=1}^{N}\frac{\nabla_i^2}{2m_i} + \sum_{i<j} V_\text{MRS}(r_{ij}),
\end{equation}
where the sum runs over the $N$ heavy (anti)quarks, $m_i$ are the heavy quark masses, and $V_\text{MRS}(r_{ij})$ is the two-body MRS potential Eq.~\eqref{eq:V_MRS} evaluated in the appropriate color representation for each pair.

\begin{figure*}
	\centering
    \includegraphics[width=0.48\linewidth]{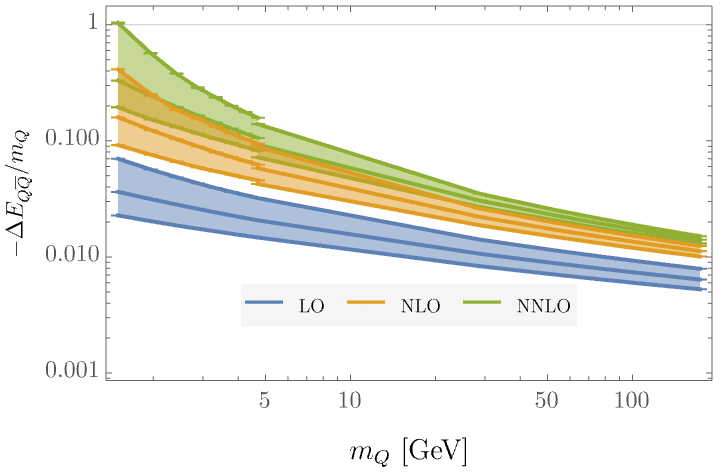} \hfill
    \includegraphics[width=0.48\linewidth]{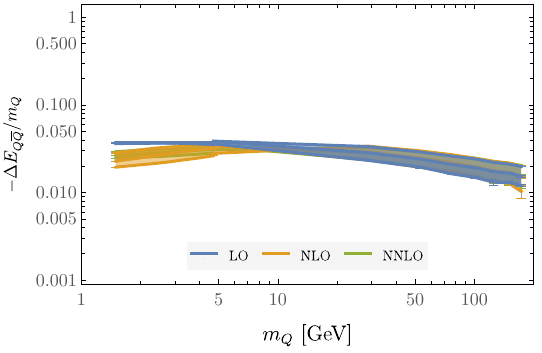}
    \caption{Quarkonium binding energies $|\Delta E_{Q\bar{Q}}|$ versus quark mass $m_Q$ 
    on a log-log scale, comparing fixed-order (left) and MRS (right) schemes at LO, NLO, 
    and NNLO. Shaded bands indicate scale variation: $\mu \in [\mu_p/2, 2\mu_p]$ for fixed-order and $\lambda \in \{1/2, 1, 2\}$ for MRS (see Sec.~\ref{app:mrs_impl}).
    Fixed-order results exhibit large order-by-order shifts and broad, non-overlapping
    bands characteristic of renormalon contamination in the static potential.
    MRS bands are narrower than fixed-order across all orders, by a factor of $\sim 2$--3 at the bottom mass and growing at lower $m_Q$, demonstrating successful renormalon subtraction and improved convergence of the perturbative series.}
	\label{fig:DeltaEQQbar_scaling}
\end{figure*}
We determine heavy quark masses by tuning $m_c$ and $m_b$ to reproduce 
spin-averaged $1S$ quarkonium binding energies within the MRS-pNRQCD framework. 
These extracted masses then serve as fixed inputs for all subsequent baryon 
and tetraquark predictions, ensuring a consistent parameter set across the 
heavy-hadron spectrum.

The quark mass $m_Q$ is tuned at each perturbative order (LO, NLO, NNLO) to
reproduce the experimentally measured spin-averaged $1S$ masses,
\begin{equation}
M_{c\bar{c}} = \frac{1}{4}(3M_{J/\psi} + M_{\eta_c}) = 3.06865(10),\text{GeV},
\end{equation}
\begin{equation}
M_{b\bar{b}} = \frac{1}{4}(3M_{\Upsilon} + M_{\eta_b}) = 9.44498(60),\text{GeV},
\end{equation}
where the experimental values are taken from Ref.~\cite{ParticleDataGroup:2024cfk}. Spin-averaging removes $\mathcal{O}(1/m_Q^2)$ hyperfine contributions, consistent with our static potential approximation at $\mathcal{O}(m_Q^0)$. 

Figure~\ref{fig:DeltaEQQbar_scaling} shows the resulting quarkonium binding energies as functions of the quark mass on a log-log scale. The figure compares fixed-order (left) and MRS (right) results across the three perturbative orders.
The MRS bands in Fig.~\ref{fig:DeltaEQQbar_scaling} are narrower than the fixed-order bands by a factor of $\sim 2$--3 at $m_Q \sim m_b$, growing at lower $m_Q$ where renormalon effects dominate. The dramatic LO collapse reflects a stationary point of $M(\mu)$ near the central scale rather than a genuine reduction in truncation error.

\begin{table*}
\caption{Spin-averaged $1S$ heavy quarkonium masses at LO, NLO, and NNLO in fixed-order (FO) and MRS schemes. For charmonium and bottomonium, quark masses are tuned to reproduce experimental values from Ref.~\cite{ParticleDataGroup:2024cfk}. Uncertainties combine GFMC statistical errors and fitting systematics; LO results are numerically exact in the FO case. For top quarkonium, $m_t = 172.52$~GeV is adopted and binding energy computed as prediction. The theory prediction of the top quarkonium comes from Ref.~\cite{maltoni2024quantumdetectionnewphysics}. MRS quark masses vary by $<1\%$ across orders vs.\ $\sim$4--11\% in FO, demonstrating successful renormalon resummation.}
\label{tab:mesonmass}
\begin{ruledtabular}
\begin{tabular}{cccccccc}
 $1S$ mesons & Order & $\alpha_s(\mu)$ & $m_Q$ (FO) [GeV]  & $M_{Q\bar{Q}}$ (FO) [GeV] & $m_Q$ (MRS) [GeV]  &   $M_{Q\bar{Q}}$ (MRS) [GeV] & $M_{Q\bar{Q}}$~\cite{ParticleDataGroup:2024cfk,maltoni2024quantumdetectionnewphysics} [GeV] \\
 \hline
  $(J/\psi,\eta_c)$ & LO  & 0.283 & 1.562 & 3.0686 & 1.56300 & 3.06865(2)&  3.06865(10)~\cite{ParticleDataGroup:2024cfk} \\
$(J/\psi,\eta_c)$ & NLO & 0.307 & 1.656  & 3.0686(7) & 1.55224 & 3.06865(13) & 3.06865(10)~\cite{ParticleDataGroup:2024cfk}\\
$(J/\psi,\eta_c)$ & NNLO & 0.291 & 1.740  & 3.0688(13) & 1.55489 & 3.06864(4)& 3.06865(10)~\cite{ParticleDataGroup:2024cfk} \\
  $(\Upsilon,\eta_b)$ & LO & 0.215 & 4.772 & 9.4452 & 4.81412  &  9.44498(11)& 9.44498(60)~\cite{ParticleDataGroup:2024cfk}   \\
 $(\Upsilon,\eta_b)$ & NLO & 0.224 & 4.863 & 9.4448(5) & 4.80312 &  9.44498(9)& 9.44498(60)~\cite{ParticleDataGroup:2024cfk} \\
 $(\Upsilon,\eta_b)$ & NNLO  & 0.222 & 4.962 & 9.4447(4)  & 4.80221 & 9.44497(8)& 9.44498(60)~\cite{ParticleDataGroup:2024cfk} \\
 $(\eta_t)$ & LO &  0.120  &  172.52  &  343.935  & 172.52  &  342.49(5)(75) & $\approx$ 343~\cite{maltoni2024quantumdetectionnewphysics} \\
 $(\eta_t)$ & NLO &  0.120  &  172.52  &  343.112(3)  & 172.52 & 342.49(8)(76) & $\approx$ 343~\cite{maltoni2024quantumdetectionnewphysics} \\
 $(\eta_t)$ & NNLO &  0.120  &  172.52  &  342.628(6)  & 172.52 & 342.47(4)(77) & $\approx$ 343~\cite{maltoni2024quantumdetectionnewphysics} \\
\end{tabular}
\end{ruledtabular}
\end{table*}

Table~\ref{tab:mesonmass} presents results for both fixed-order (FO) and MRS schemes. In the FO scheme, tuned quark masses increase substantially with perturbative order: $m_c$ increases from 1.56~GeV at LO to 1.74~GeV at NNLO (11\% variation), while $m_b$ increases from 4.77~GeV to 4.96~GeV (4\% variation). This reflects the $\mu$ dependence and renormalon-induced instabilities characteristic of fixed-order perturbation theory. In contrast, MRS-tuned masses decrease mildly with order and stabilize: $m_c$ ranges from 1.56~GeV to 1.55~GeV (sub-percent variation), and $m_b$ from 4.81~GeV to 4.80~GeV (sub-percent variation). This dramatic reduction in order-by-order scatter demonstrates that MRS successfully isolates and resums the leading renormalon.
The extracted masses are free parameters defined by $2m_Q + E_\text{GFMC}(m_Q) = M_\text{exp}$, where the nonperturbative power correction (Sec.~\ref{app:mrs_impl}) is implicitly absorbed into $m_Q$. The purpose of this tuning is not to determine quark masses but to fix the one free parameter per flavor, so that predictions for mixed-flavor mesons, baryons, and tetraquarks are parameter free. Since $m_Q$ is tuned, replacing it by another short-distance mass definition (e.g., the 1S or PS mass) is a reparametrization absorbed by the tuning, and leaves our predictions for other states unchanged up to effects of the order of the neglected $1/m_Q$ corrections. In principle, quark masses from MRS analyses in lattice QCD~\cite{Brambilla:2017hcq, FermilabLattice:2018est} could be used as inputs instead, making even the equal-flavor meson sector a prediction. This would require matching the implementation details of the MRS scheme (the $\mu'(r)$ prescription, the treatment of the power correction, and the perturbative truncation order) between the two approaches, and is left to future work.

In the fixed-order scheme, conventional scale variation $\mu \in [\mu_p/2, 2\mu_p]$ produces $\sim 100\,\text{MeV}$ % getting rid of abuse of "big O" notation
shifts in NNLO bottomonium binding energies (less for top, more for charm).
In the MRS scheme, the analogous variation $\lambda \in \{1/2, 1, 2\}$ (Sec.~\ref{app:mrs_impl}) reduces these shifts by a factor of $\sim 2$--3 in the bottom region, with larger reductions at lower $m_Q$.
MRS scale dependence enters through the dynamical scale $\mu'(r)$ in the Borel-resummed potential~\eqref{eq:V_MRS}, varying with the short-distance physics probed at each separation rather than being set by a fixed renormalization point. The reduced MRS band reflects suppression of renormalon-induced scale dependence; it is a diagnostic of scheme stability rather than a quantitative truncation-error estimate. The LO band's apparent collapse reflects a stationary point of $M(\mu)$ near the central scale, not a meaningful reduction in truncation uncertainty.

Although the top quark decays before hadronization, near-threshold $t\bar{t}$ production at the LHC is sensitive
to Coulombic bound-state effects that modify differential cross sections and spin
correlations~\cite{Fuks:2021xje,Fuks:2025toponiumSimulation,Garzelli:2024uhe}.
These analyses employ NRQCD Green's functions to capture the dynamics of the
color-singlet pseudoscalar $\eta_t$ state -- precisely the regime described by our
pNRQCD framework.
Threshold predictions for $t\bar{t}$ production at the LHC~\cite{Fuks:2021xje,Fuks:2025toponiumSimulation,Garzelli:2025toponiumLHC,Garzelli:2024uhe} incorporate the top quark width using fixed-order potentials in the NRQCD Green's functions; incorporating the MRS-resummed potential in these Green's functions would provide improved perturbative stability.
We observe mild order-by-order variation ($\sim$20~MeV between LO and NNLO in MRS) in $M_{t\bar{t}}$ predictions, which reflects that $\alpha_s(m_t) \approx 0.108$ is small and makes top quarkonium particularly amenable to perturbative treatment even before MRS resummation.
For the top quark, we adopt $m_t = 172.52$~GeV from a combination of six ATLAS and nine CMS measurements~\cite{Faltermann:2024fzb} and predict $M_{t\bar{t}} = 342.47(4)(77)$~GeV at NNLO with MRS in the zero-width limit; see Table~\ref{tab:mesonmass}.

\subsection{Comparison with other \texorpdfstring{\boldmath$m_Q$}{mQ} determinations}

The MRS quark masses extracted from lattice QCD by the Fermilab Lattice, MILC, and TUMQCD collaborations~\cite{FermilabLattice:2018est} are $m_{c,{\rm MRS}} = 1.392(11)$~GeV and $m_{b,{\rm MRS}} = 4.749(18)$~GeV.
Our NNLO MRS values from Table~\ref{tab:mesonmass} are $m_c = 1.555$~GeV and $m_b = 4.802$~GeV.
The bottom mass differs from the lattice-QCD determination by 53~MeV, and the charm mass by 163~MeV.
Both discrepancies are larger than a naive estimate of $\mathcal{O}(1/m_Q^2)$ relativistic corrections ($\sim$2.5~MeV for bottom, $\sim$22~MeV for charm from $\alpha_s^2$ shifts to the binding energy).
The differences likely reflect a combination of such corrections, residual scheme-convention differences in the MRS power-correction subtraction, and $1/m_Q$ corrections not yet included in our calculation.
Note also that the lattice-QCD determination~\cite{FermilabLattice:2018est} was obtained from computing the heavy-light pseudoscalar meson mass and, thus, applied MRS to the pole-mass--$\overline{\rm MS}$-mass relation, not the static potential.

\subsection{Mixed-flavor systems}

\begin{table*} % [tb]
\caption{MRS predictions for $1S$ spin-averaged mixed-flavor heavy meson masses using
quark masses from Table~\ref{tab:mesonmass}.
The first parenthetical is the GFMC statistical error; the second is the maximum deviation from the central scale under the MRS scale variation $\lambda \in \{1/2, 1, 2\}$, both in units of the last digit. The scale-variation range probes residual $\mu$-dependence and serves as a diagnostic of scheme stability rather than a quantitative truncation-error estimate.}
\label{tab:mixed_mesonmass}
\begin{ruledtabular}
\begin{tabular}{ccccc}
    $1S$ mesons & Order & $m_{Q_1 Q_2}$ [GeV] & $M_{Q_1\bar{Q}_2}$ [GeV] & Predicted $\bar{M}_{Q_1\bar{Q}_2}$ [GeV] \\
    \hline
    $b\bar{c}$ & LO & 2.35983 & 6.28978(3)(132)  & 6.316(7)~\cite{ParticleDataGroup:2024cfk,Gregory:2009hq} \\
    $b\bar{c}$ & NLO  & 2.34624 & 6.29386(8)(1345)  & 6.316(7)~\cite{ParticleDataGroup:2024cfk,Gregory:2009hq}  \\
    $b\bar{c}$ & NNLO  & 2.34916 & 6.29121(4)(731) & 6.316(7)~\cite{ParticleDataGroup:2024cfk,Gregory:2009hq} \\
    $t\bar{c}$ & LO  & 3.09793 & 173.96806(4)(217) &  --  \\
    $t\bar{c}$ & NLO  & 3.07680 & 173.98431(6)(1707) &  --   \\
    $t\bar{c}$ & NNLO  & 3.08200 & 173.98462(4)(883) &  --  \\
    $t\bar{b}$ & LO  & 9.36687 & 176.99884(40)(4112)  &  --  \\
    $t\bar{b}$ & NLO  & 9.34603  & 176.99480(32)(2881) &   --  \\
    $t\bar{b}$ & NNLO  & 9.34432 & 177.00684(39)(2372) &  --  \\
\end{tabular}
\end{ruledtabular}
\end{table*}

For mixed-flavor mesons such as $bc$, $ct$, and $bt$, we use the MRS-tuned quark masses from Table~\ref{tab:mesonmass} as inputs to the GFMC.
The reduced mass for unequal-mass systems is
\begin{equation}
    m_{Q_1\cdots Q_N} = \frac{N}{\displaystyle\sum_{i=1}^{N} \frac{1}{m_{Q_i}}},
\end{equation}
Table~\ref{tab:mixed_mesonmass} presents predictions for the spin-averaged meson masses, $\bar{M}_{Q_1\bar{Q}_2} = \frac{1}{4}(M_{Q_1\bar{Q}_2}(1^1S_0) + 3M_{Q_1\bar{Q}_2}(1^3S_1))$.
The spin-averaged $B_c$ meson mass prediction of $6.29121(4)(731)$~GeV at NNLO MRS lies $\sim$25~MeV below the value $\bar{M}_{B_c} = 6.316(7)$~GeV, constructed from the experimental $M_{B_c}=6.2745(3)$~GeV~\cite{ParticleDataGroup:2024cfk} and the lattice-QCD $M_{B_c^*}=6.330(9)$~GeV~\cite{Gregory:2009hq}.
The difference is comparable to the MRS scale-variation range.
For $ct$ and $bt$ systems, no experimental data exist, but our predictions provide benchmarks for future measurements or lattice QCD comparisons.

\section{Triply-heavy baryons}\label{sec:qcd_baryons}

With the heavy-quark masses $m_c$ and $m_b$ fixed from quarkonium 
(Table~\ref{tab:mesonmass}), pNRQCD provides predictions for the spectrum of 
triply-heavy baryons. The static Hamiltonian consists of pairwise 
color-antisymmetric quark--quark interactions plus three-body forces that 
first appear at NNLO~\cite{Brambilla:2004jw,Brambilla:1999qa,Brambilla:2006wp,Assi:2023cfo}. 
At $\mathcal{O}(m_Q^0)$, the two-body potential acting on a color-singlet 
baryon state $\ket{B}$ is
\begin{align}
    \hat V^{\psi\psi}\ket B &= \sum_{I<J}V^{\psi\psi}_{\rm A}(r_{IJ})\ket B \nonumber \\ &=
        -\sum_{I<J}\frac{C_B\alpha_V^{\rm MRS}(r_{IJ})}{r_{IJ}}\ket B,
\end{align}
where $C_B = C_F/2 = 2/3$ for QCD and the sum runs over the three quark pairs.
Three-body contributions appear at $\mathcal{O}(\alpha_s^3)$, suppressed by 
$\alpha_s^2$ relative to the leading Coulomb interaction; in our previous 
fixed-order analysis~\cite{Assi:2023cfo} they modified binding energies by less 
than one percent for $\alpha_s \lesssim 0.3$. In the MRS scheme, the two-body potential receives the full resummation of the leading renormalon; three-body potentials, which first appear at NNLO, are included at fixed order since no resummation formula is available for them. 

The dominance of pairwise Coulombic interactions motivates a trial wavefunction 
built as a product of two-body Coulomb ground states,
\begin{equation}
        \Psi_T^{QQQ}(\bs{R}) = \prod_{I=1}^{3} \prod_{J<I} \frac{1}{\sqrt{\pi}a_0^{3/2}} e^{-|\bs{r}_{IJ}|/a_0},
\label{eq:baryon_trial}
\end{equation}
$\bs{R}=(\bs{r}_1,\bs{r}_2,\bs{r}_{3})$.
As in the quarkonium case, the Bohr radius is set by evaluating the MRS-resummed potential at $r^*$:
\begin{equation}
  a_0 = \frac{2}{C_B\, \alpha_V^{\rm MRS}(r^*)\, m_Q},
\end{equation}
with $C_F\to C_B$ relative to Eq.~\eqref{eq:aL_trial}. In our previous fixed-order study~\cite{Assi:2023cfo}, the simple product of $n=1$, $l=0$ Coulomb factors yielded variational bounds within a percent of more elaborate ansätze. We adopt the same trial wavefunction form here.

For mixed-flavor baryons, we use the corresponding MRS-tuned quark masses from Table~\ref{tab:mesonmass} as inputs to the GFMC.
The renormalization scale enters through the MRS-resummed potential~\eqref{eq:V_MRS} via the dynamical scale $\mu'(r)$, with residual scale dependence probed by rescaling $\lambda \in \{1/2, 1, 2\}$ as described in Sec.~\ref{app:mrs_impl}.

\subsection{Baryon binding energies and \texorpdfstring{$m_Q$}{mQ} scaling}

\begin{figure*} % [!t]
	\centering
    \includegraphics[width=0.48\linewidth]{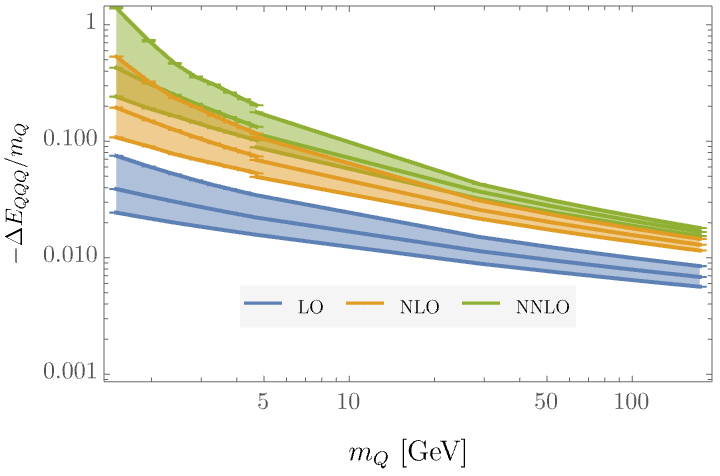} \hfill
    \includegraphics[width=0.48\linewidth]{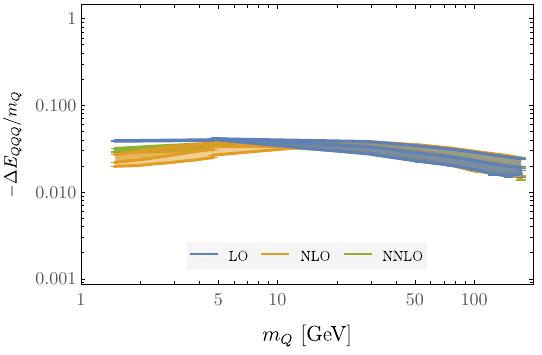}
	\caption{Triply-heavy baryon binding energies $|\Delta E_{QQQ}|$ versus quark mass $m_Q$ on a log-log scale, comparing fixed-order (left) and MRS (right) schemes at LO, NLO, and NNLO.
    Shaded bands indicate scale variation: $\mu \in [\mu_p/2, 2\mu_p]$ for fixed-order and $\lambda \in \{1/2, 1, 2\}$ for MRS (see Sec.~\ref{app:mrs_impl}).
    Fixed-order results exhibit large order-by-order shifts and broad scale-variation bands,
    reflecting renormalon sensitivity in the static potential. MRS reorganization yields
    overlapping bands across perturbative orders and reduces scale dependence,
    demonstrating improved convergence of the perturbative series.}
	\label{fig:DeltaEQQQ_scaling}
\end{figure*}
Figure~\ref{fig:DeltaEQQQ_scaling} shows triply-heavy baryon binding energies as functions of quark mass on a log-log scale, comparing fixed-order (left) and MRS (right) results at LO, NLO, and NNLO. In the fixed-order scheme, the three orders are well separated with no overlap and scale-variation bands spanning factors of 2--3 in the binding energy. The MRS scheme brings the three orders into partial overlap, with scale-variation bands narrowing by roughly a factor of 2--3. The improvement is most visible at low $m_Q$ where $\alpha_s$ is largest.

\begin{figure*} % [!t]
	\centering
    \includegraphics[width=0.48\linewidth]{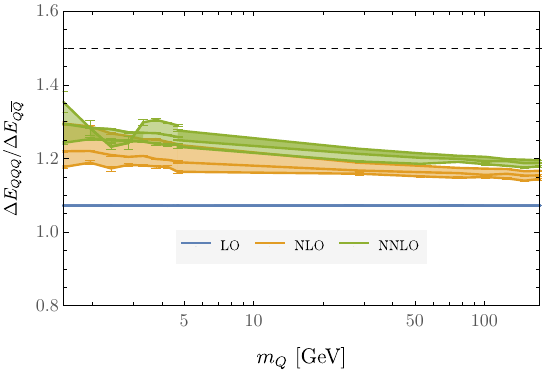} \hfill
    \includegraphics[width=0.48\linewidth]{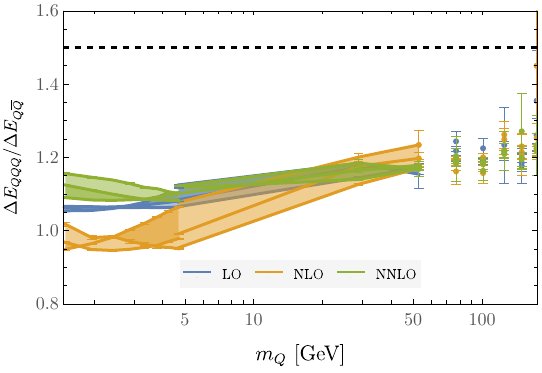}
	\caption{Binding energy ratio $\Delta E_{QQQ}/\Delta E_{Q\bar{Q}}$ versus $m_Q$ with scale
variation ($\mu \in [\mu_p/2, 2\mu_p]$ for FO, $\lambda \in \{1/2, 1, 2\}$ for MRS), comparing fixed-order (left) and MRS (right) schemes
at LO, NLO, and NNLO. The dashed line indicates the Detmold bound 
$\Delta E_{QQQ}/\Delta E_{Q\bar{Q}} \leq 3/2$~\cite{Detmold:2014iha}, which follows from 
QCD positivity constraints on meson-meson interactions. MRS results exhibit well-behaved perturbative convergence with 
overlapping scale-variation bands across orders, while fixed-order results show large 
gaps between LO and higher orders characteristic of renormalon contamination in the 
static potential.}
	\label{fig:baryon_meson_ratio}
\end{figure*}

Figure~\ref{fig:baryon_meson_ratio} shows the baryon-to-meson binding-energy ratio $\Delta E_{QQQ}/\Delta E_{Q\bar{Q}}$ as a function of $m_Q$. In the fixed-order scheme, LO is a flat line near 1.08, while NLO and NNLO lie higher at $\sim$1.15--1.35 with significant scale-variation bands. NNLO is closer to NLO than either is to LO.
In the MRS scheme, all three orders overlap across the full $m_Q$ range, with a spread of roughly 0.9--1.2 at charm masses narrowing toward higher $m_Q$. At high masses ($m_Q \gtrsim 50$~GeV), statistical fluctuations in the ratio become visible. The difference between the two schemes is most pronounced at low $m_Q$, where $\alpha_s$ is largest and renormalon effects in the individual binding energies are most significant.

\subsection{Comparison with lattice QCD}

Table~\ref{tab:baryonmass} compares our predictions with lattice-NRQCD results~\cite{Brown:2014ena,Meinel:2010pw,Mathur:2018epb, Mathur:2022nez}, which include $\mathcal{O}(1/m_Q^2)$ relativistic corrections.
Both fixed-order and MRS NNLO results undershoot lattice-QCD masses across all four baryon species. Fixed-order NNLO lies 80--110~MeV below lattice QCD, while MRS NNLO lies 125--175~MeV below -- further from lattice QCD in absolute terms, reflecting the shift in central values that accompanies renormalon subtraction. The advantage of MRS is not in the central value but in the reduced scale variation ($\sim 10$~MeV versus $\sim 800$~MeV), which makes the discrepancy with lattice QCD unambiguous and interpretable as missing $\mathcal{O}(1/m_Q)$ velocity-dependent and $\mathcal{O}(1/m_Q^2)$ spin-dependent corrections rather than perturbative instability.

\begin{table*} % [tb]
\caption{Comparison of triply-heavy baryon mass results with other pNRQCD and lattice QCD calculations. All masses in GeV, obtained using $\alpha_s$ and $m_Q$ from Table~\ref{tab:mesonmass}. For both ``Previous FO'' and ``This work MRS'' columns, the first parenthetical is the GFMC statistical error; the second is the maximum deviation from the central scale under the scale variation ($\mu \in [\mu_p/2, 2\mu_p]$ for FO, $\lambda \in \{1/2, 1, 2\}$ for MRS), both in units of the last digit. The scale-variation range probes residual $\mu$-dependence and serves as a diagnostic of scheme stability rather than a quantitative truncation-error estimate. Lattice-QCD uncertainties are statistical and systematic respectively.}
\label{tab:baryonmass}
\begin{ruledtabular}
\begin{tabular}{llcccc}
Baryon & Order & This work FO~\cite{Assi:2023cfo} & This work MRS & Variational & Lattice QCD \\
\hline
$\Omega_{ccc}$ & LO   & 4.627(0)(58)   & 4.62808(3)(106)   & 4.76(6)~\cite{Jia:2006gw} & 4.796(8)(18)~\cite{Brown:2014ena} \\
               & NLO  & 4.646(1)(331)  & 4.62218(11)(768)  & & \\
               & NNLO & 4.690(1)(849) & 4.61855(5)(344)   & 4.97(20)~\cite{Llanes-Estrada:2011gwu} & \\
\hline
$\Omega_{ccb}$ & LO   & 7.829(0)(60)   & 7.86102(4)(124)   & 7.98(7)~\cite{Jia:2006gw} & 8.007(9)(20)~\cite{Brown:2014ena} \\
               & NLO  & 7.871(1)(327)  & 7.86011(8)(1203)  & & 8.005(6)(11)~\cite{Mathur:2018epb} \\
               & NNLO & 7.910(1)(779)  & 7.85108(6)(524)   & 8.20(15)~\cite{Llanes-Estrada:2011gwu} & \\
\hline
$\Omega_{cbb}$ & LO   & 11.026(0)(66)  & 11.07877(4)(192)  & 11.48(12)~\cite{Jia:2006gw} & 11.195(8)(20)~\cite{Brown:2014ena} \\
               & NLO  & 11.061(1)(328) & 11.08393(6)(1894) & & 11.194(5)(12)~\cite{Mathur:2018epb} \\
               & NNLO & 11.116(1)(753) & 11.07011(6)(827) & 11.34(26)~\cite{Llanes-Estrada:2011gwu} & \\
\hline
$\Omega_{bbb}$ & LO   & 14.210(0)(76)  & 14.24318(7)(630)  & 14.76(18)~\cite{Jia:2006gw} & 14.371(4)(12)~\cite{Meinel:2010pw} \\
               & NLO  & 14.249(1)(338) & 14.24938(13)(3187) & & 14.366(9)(20)~\cite{Brown:2014ena} \\
               & NNLO & 14.282(1)(706) & 14.23110(7)(1116) & 14.57(25)~\cite{Llanes-Estrada:2011gwu} & 14.366(7)(9)~\cite{Mathur:2022nez} \\
\end{tabular}
\end{ruledtabular}
\end{table*}

\begin{table*} % [tb]
\caption{MRS predictions for triply-heavy baryon masses containing top quarks.
All masses in GeV. The first parenthetical is the GFMC statistical error; the second is the maximum deviation from the central scale under the MRS scale variation $\lambda \in \{1/2, 1, 2\}$, both in units of the last digit. The scale-variation range probes residual $\mu$-dependence and serves as a diagnostic of scheme stability rather than a quantitative truncation-error estimate.
These states cannot form due to top decay but serve as QCD benchmarks
in the $m_Q \to \infty$ limit.}
\label{tab:topbaryonmass}
\begin{ruledtabular}
\begin{tabular}{l c c c}
Baryon & LO & NLO & NNLO \\
\hline
 $\Omega_{tcc}$ & 175.55412(4)(147) & 175.56716(7)(1488) & 175.55827(6)(648) \\
 $\Omega_{tcb}$ & 178.75739(6)(234) & 178.77631(6)(2260) & 178.76384(5)(995) \\
 $\Omega_{tbb}$ & 181.85680(11)(1935) & 181.86651(20)(4431) & 181.85756(8)(1444) \\
 $\Omega_{ttc}$ & 346.41934(8)(342) & 346.45037(8)(2922) & 346.44124(7)(1273) \\
 $\Omega_{ttb}$ & 349.32836(41)(6690) & 349.32326(37)(5008) & 349.33347(34)(4681) \\
 $\Omega_{ttt}$ & 514.45(7)(86) & 514.46(5)(95) & 514.46(4)(94) \\
\end{tabular}
\end{ruledtabular}
\end{table*}

Previous pNRQCD variational studies obtained masses well above lattice QCD: Jia~\cite{Jia:2006gw} using LO potentials and Llanes-Estrada \emph{et al.}~\cite{Llanes-Estrada:2011gwu} using full NNLO potentials. Our previous fixed-order results~\cite{Assi:2023cfo} are shown in Table~\ref{tab:baryonmass}.
Our NNLO+MRS results undershoot lattice QCD by 125--175~MeV. Order-by-order shifts stay below $\sim$20~MeV in MRS, compared with $\sim$20--55~MeV in fixed order.

Table~\ref{tab:topbaryonmass} extends these predictions to triply-heavy baryons 
containing top quarks. The top quark decays via $t \to Wb$ before hadronization, 
so these states cannot exist as asymptotic hadrons. However, the recent observation 
of threshold enhancements in $t\bar{t}$ production at the LHC~\cite{CMS:2025kzt} 
demonstrates that Coulombic bound-state effects can produce observable signatures 
even when constituent lifetimes preclude true hadron formation. For triply-topped 
systems, production cross sections are highly suppressed, but the same near-threshold 
dynamics could in principle affect multi-top final states at future colliders.

Other recent predictions for top-containing baryons use QCD sum rules or potential models with Higgs-exchange and nonperturbative condensates~\cite{Zhu:2025ttt,Najjar:2025ttt,Shekari:2026ttb}; the same MRS methodology extends to top-containing tetraquarks (Sec.~\ref{sec:qcd_tetras}).

\begin{figure*} % [t!]
    \centering
    \includegraphics[width=0.48\linewidth]{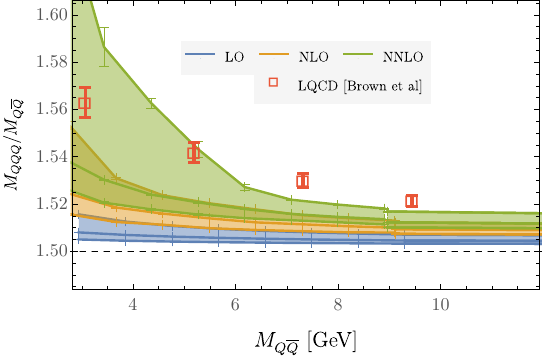} \hfill
    \includegraphics[width=0.48\linewidth]{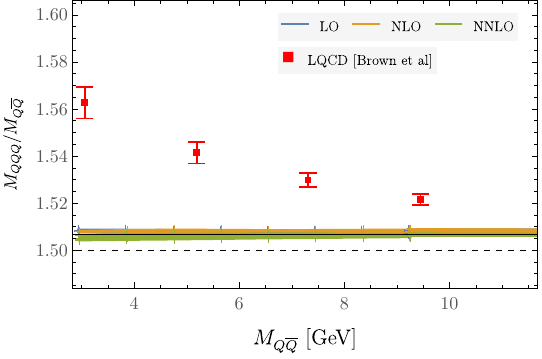}
    \caption{Baryon-to-meson mass ratio $M_{QQQ}/M_{Q\bar{Q}}$ versus meson mass $M_{Q\bar{Q}}$, comparing fixed-order (left) and MRS (right) schemes at LO, NLO, and NNLO with lattice-QCD results~\cite{Brown:2014ena} (red points).
    The dashed line shows the Detmold bound $M_{QQQ}/M_{Q\bar{Q}} \geq 3/2$~\cite{Detmold:2014iha}. 
    Fixed-order calculations exhibit limited convergence and pronounced renormalization-scale sensitivity, particularly at NNLO. 
    The MRS scheme significantly reduces this scale dependence, yielding tightly overlapping bands across perturbative orders. 
    The apparent consistency of fixed-order results with lattice QCD primarily arises from their enlarged scale-variation bands. 
    Overall, the lattice data show a systematic tendency toward the predictions in the heavier-mass regime, where the heavy-quark expansion is expected to be more robust.}
    \label{fig:boundQCD}
\end{figure*}

Figure~\ref{fig:boundQCD} compares baryon-to-meson mass ratios $M_{QQQ}/M_{Q\bar{Q}}$ with lattice-QCD benchmarks. Both panels confirm the Detmold inequality $M_{QQQ} \geq (3/2)M_{Q\bar{Q}}$~\cite{Detmold:2014iha}. In the fixed-order scheme (left), the NNLO band is broad enough to overlap with some lattice-QCD points. In the MRS scheme (right), scale variation is dramatically reduced and the three perturbative orders cluster tightly between 1.502 and 1.508. The lattice-QCD data points lie systematically higher at $\sim$1.52--1.57, consistent with the 125--175~MeV mass discrepancy identified in Table~\ref{tab:baryonmass} and attributable to missing $\mathcal{O}(1/m_Q)$ corrections shifting the ratio.

The proximity of the MRS ratios to $3/2$ has a simple origin: $M_{QQQ}/M_{Q\bar{Q}} - 3/2 = (\Delta E_{QQQ} - \tfrac{3}{2}\Delta E_{Q\bar{Q}})/M_{Q\bar{Q}}$, so the quark masses cancel and the deviation is set by binding energies. In the MRS scheme these are small (e.g., $|\Delta E_{c\bar{c}}| \approx 40$~MeV), since the renormalon-related constant is absorbed into $m_Q$. The lattice-QCD ratios instead correspond to $M_{QQQ} - \tfrac{3}{2}M_{Q\bar{Q}} \approx 200$~MeV, roughly independent of flavor. Reproducing this requires effects beyond our static perturbative potentials, such as $1/m_Q$ corrections, spin-dependent interactions (the $\Omega_{ccc}$ and $\Omega_{bbb}$ ground states have $J=3/2$), and power corrections in the non-singlet channels, which we neglect.

The mass ratio $M_{QQQ}/M_{Q\bar{Q}}$ partially cancels the leading renormalon between numerator and denominator. In the fixed-order scheme, scale-variation bands grow with perturbative order: $\sim$1\% at LO, $\sim$2\% at NLO, and $\sim$10\% at NNLO near $m_Q \sim m_c$, reflecting renormalon contamination in the individual masses. In the MRS scheme, all three orders fall within $\sim$0.3\% across the full mass range and overlap closely, as seen in Fig.~\ref{fig:boundQCD}.

\section{All-heavy tetraquarks}
\label{sec:qcd_tetras}

In this section, we extend the MRS-pNRQCD framework to fully heavy tetraquarks $QQ\bar{Q}'\bar{Q}'$. We compute the binding energy ratio as a function of the mass ratio $m_Q/m_{Q'}$ and identify bound configurations for physical quark masses.

Previous studies using variational Monte Carlo at leading order~\cite{Czarnecki:2017vco} 
and Green's function Monte Carlo at next-to-leading order~\cite{Assi:2023dlu} 
established that equal-mass configurations such as $bb\bar{b}\bar{b}$ or $cc\bar{c}\bar{c}$ 
do not form bound states\footnote{Using single-gluon-exchange potentials, Ref.~\cite{Anwar:2017toa} reaches different conclusions regarding the existence of equal-mass fully-heavy tetraquark bound states.}: the attractive $\mathbf{3}\otimes\bar{\mathbf{3}}$ diquark-antidiquark 
channel cannot overcome the repulsive contributions present when all constituents share the same mass. 
Bound tetraquarks exist only in configurations of the form $QQ\bar{Q}'\bar{Q}'$ with a 
sufficiently large mass hierarchy $m_Q \ll m_{Q'}$. Both Ref.~\cite{Czarnecki:2017vco} and 
Ref.~\cite{Assi:2023dlu} independently identified a critical mass ratio 
$m_Q/m_{Q'} \lesssim 0.15$ at LO, with Ref.~\cite{Assi:2023dlu} finding a slightly 
lower threshold $m_Q/m_{Q'} \lesssim 0.12$ at NLO.

Following Refs.~\cite{Hylleraas:1928,Assi:2023dlu}, we employ 
Hylleraas-type molecular trial wavefunctions of the form
\begin{align}
  \Psi_H(\bs{R}; a, b, c) \propto {}&
  e^{-|\bs{r}_{13}|/a}\, e^{-|\bs{r}_{24}|/a}\, 
  e^{-|\bs{r}_{14}|/b}\, e^{-|\bs{r}_{23}|/b}\, 
  \nonumber\\&\times
  e^{-|\bs{r}_{12}|/c}\, e^{-|\bs{r}_{34}|/c},
  \label{eq:tetra_hylleraas}
\end{align}
where $a$, $b$, and $c$ are variational parameters encoding correlations at different length scales: 
$a$ governs the tightly-bound quark-antiquark pairs, $b$ describes weaker inter-pair correlations 
between opposite-type constituents, and $c$ captures same-type correlations. 
These spatial wavefunctions are combined with color tensors projecting onto 
$\mathbf{3}\otimes\bar{\mathbf{3}}$ or $\mathbf{6}\otimes\bar{\mathbf{6}}$ configurations. 
Using the same quark masses tuned to quarkonium (Table~\ref{tab:mesonmass}), we extend our 
GFMC analysis to these four-quark states.

The tetraquark Hamiltonian includes two-body color-singlet and color-octet $Q\bar{Q}$ as well as color-antisymmetric and color-symmetric $QQ$ potentials, all of which receive MRS resummation as in the meson and baryon sectors. Three-body
potentials involving $QQ\bar{Q}$ and $Q\bar{Q}\bar{Q}$ configurations and
four-body potentials first appear at NNLO~\cite{Assi:2023dlu}. Our tetraquark
results are presented at LO and NLO only; extending to NNLO with these
multi-body potentials is left for future work.

\subsection{Binding energy ratios and mass hierarchy}

\begin{figure*} % [t!]
    \centering
    \includegraphics[width=0.48\linewidth]{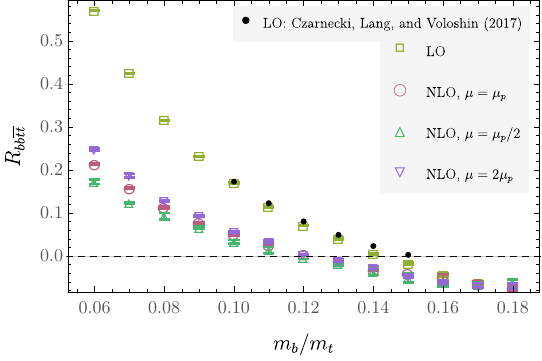} \hfill
    \includegraphics[width=0.48\linewidth]{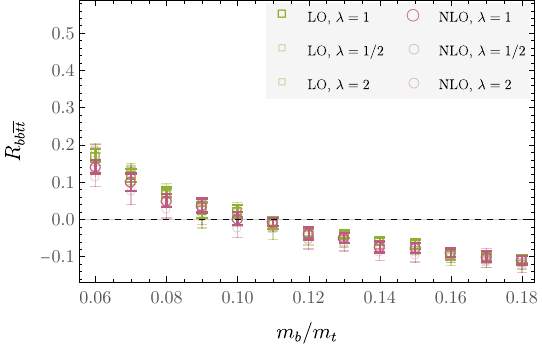}
    \caption{Binding energy ratios $R_{QQ\bar{Q}'\bar{Q}'}$ versus heavy-light mass ratio $m_Q/m_{Q'}$, comparing fixed-order (left) and MRS (right) schemes.
    Distinct marker styles denote scale choices: $\mu \in \{\mu_p/2, \mu_p, 2\mu_p\}$ for FO and $\lambda \in \{1/2, 1, 2\}$ for MRS.
    MRS requires a smaller mass ratio for binding ($m_Q/m_{Q'} \sim 0.10$ versus $\sim 0.12$--$0.15$ at fixed order). Error bars denote GFMC statistical uncertainties.}
    \label{fig:unequalmass_binding_mrs}
\end{figure*}

Figure~\ref{fig:unequalmass_binding_mrs} shows the binding energy ratio
\begin{equation}
  R_{QQ\bar{Q}'\bar{Q}'} \equiv 
  \frac{\Delta E_{QQ\bar{Q}'\bar{Q}'} - 2\,\Delta E_{Q\bar{Q}'}}{2\,\Delta E_{Q\bar{Q}'}}
  \label{eq:Rtetra}
\end{equation}
as a function of the heavy-light mass ratio $m_Q/m_{Q'}$ for fixed-order (left) and MRS (right) schemes. Positive values indicate binding below the two-meson threshold. Discrete points at three scale choices ($\mu \in \{\mu_p/2, \mu_p, 2\mu_p\}$ for FO, $\lambda \in \{1/2, 1, 2\}$ for MRS) show the scale dependence; all points use GFMC with $\mathbf{3}\otimes\bar{\mathbf{3}}$ trial wavefunctions.

In the fixed-order scheme, LO (single scale choice) gives the largest $R$ values with a binding threshold near $m_Q/m_{Q'} \simeq 0.14$--$0.15$. At NLO, the three scale choices ($\mu \in \{\mu_p/2, \mu_p, 2\mu_p\}$) yield closely grouped points with a slightly lower threshold near $\simeq 0.12$--$0.13$, consistent with Refs.~\cite{Czarnecki:2017vco,Assi:2023dlu}. In the MRS scheme, LO acquires scale dependence through $\lambda \in \{1/2, 1, 2\}$ (spread $\sim$0.02--0.03 in $R$) and the binding threshold shifts to $m_Q/m_{Q'} \simeq 0.10$--$0.11$. At NLO, MRS points cluster tightly with a threshold near $m_Q/m_{Q'} \simeq 0.10(1)$.

\subsection{Bound tetraquark states with top quarks}

For the physical heavy-quark masses tuned in Table~\ref{tab:mesonmass}, we find bound configurations in the channels $bb\bar{t}\bar{t}$, $bc\bar{t}\bar{t}$, and $cc\bar{t}\bar{t}$. Table~\ref{tab:tetra} summarizes the corresponding total masses and binding energies, comparing fixed-order and MRS schemes. Although top quarks decay before hadronization in the Standard Model, these results define well-posed QCD benchmarks and test the predictive stability of multi-quark effective field theories.

\begin{table*} % consider switching to \begin{table} since it's now narrow
\caption{Masses and binding energies for tetraquarks composed of $\{c,b,t\}$ quarks, defined relative to the lightest quarkonium-quarkonium threshold, computed in fixed-order and MRS-pNRQCD frameworks at LO and NLO. Ground-state energies determined via GFMC with $\mathbf{3}\otimes\bar{\mathbf{3}}$ Hylleraas-type trial wavefunctions. The first parenthetical is the GFMC statistical error; the second is the full range of the scale variation ($\mu \in [\mu_p/2, 2\mu_p]$ for FO, $\lambda \in \{1/2, 1, 2\}$ for MRS), in units of the last digit. The scale-variation range probes residual $\mu$-dependence and serves as a diagnostic of scheme stability rather than a quantitative truncation-error estimate. Masses use $m_t = 172.52$~GeV with $m_c$ and $m_b$ from Table~\ref{tab:mesonmass}.}
\label{tab:tetra}
\begin{ruledtabular}
\begin{tabular}{llcccc}
 & & \multicolumn{2}{c}{Fixed-Order} & \multicolumn{2}{c}{MRS} \\
\cline{3-4}\cline{5-6}
Flavor & Order & $M$~[GeV] & $B$~[MeV] & $M$~[GeV] & $B$~[MeV] \\
\hline
\multirow{2}{*}{$bb\bar{tt}$} & LO  & 354.251(1)(279)  & 518(1)(179)  & 353.496(6)(454)  & 502(6)(519)  \\
                              & NLO & 354.019(3)(204)  & 476(3)(19)   & 353.479(10)(371) & 509(10)(340) \\[2pt]
\multirow{2}{*}{$bc\bar{tt}$} & LO  & 350.853(1)(423)  & 743(1)(315)  & 350.378(11)(418) & 589(11)(437) \\
                              & NLO & 350.840(4)(102)  & 516(4)(135)  & 350.376(9)(401)  & 604(9)(410)  \\[2pt]
\multirow{2}{*}{$cc\bar{tt}$} & LO  & 347.494(1)(551)  & 995(1)(477)  & 347.210(11)(415) & 726(11)(415) \\
                              & NLO & 347.675(12)(500) & 622(12)(240) & 347.216(8)(414)  & 742(8)(479)  \\
\end{tabular}
\end{ruledtabular}
\end{table*}

The fixed-order results show substantial scale-variation bands -- $\sim$35--50\% in binding energy under $\mu \in [\mu_p/2, 2\mu_p]$ at LO, reducing to $\sim$5--40\% at NLO. Binding energies range from 476~MeV for $bb\bar{t}\bar{t}$ to 622~MeV for $cc\bar{t}\bar{t}$ at fixed-order NLO, corresponding to 0.13--0.18\% of the total tetraquark mass.

For equal or near-equal mass systems such as $bb\bar{b}\bar{b}$, $bc\bar{b}\bar{c}$, or $cc\bar{b}\bar{b}$, the effective mass extracted from GFMC remains at or slightly above the two-quarkonium threshold, showing no sign of binding within our variational space. This absence is consistent with the lack of a sufficiently deep color-Coulomb pocket once the diquark mass hierarchy is lost. The critical ratio $m_Q/m_{Q'} \sim 0.10$ (MRS) vs. $m_Q/m_{Q'} \sim 0.12$ (fixed-order) marks a sharp transition: below these thresholds, the diquark-antidiquark configuration lies below the two-meson threshold; above them, the two-meson threshold becomes the lower-energy state.

In all bound cases, the GFMC ground-state energy is lower for $\mathbf{3}\otimes\bar{\mathbf{3}}$ trial wavefunctions than for $\mathbf{6}\otimes\bar{\mathbf{6}}$, by 50--150~MeV. This is consistent with the diquark-molecule picture in which the attractive color-triplet channel dominates for sufficiently disparate masses, though the slow rate of color-angle convergence in GFMC imaginary-time evolution makes exploring a wide range of initial-state color configurations essential and warrants further study.

At fixed-order NLO, the bound configurations have binding energies of 400--600~MeV below the two-meson threshold, with scale variation of 20--50\%.

\section{Conclusions}
\label{sec:out}

The central advance of this work is the systematic implementation of minimal renormalon subtraction (MRS) within the pNRQCD-QMC framework. By removing the leading renormalon ambiguity from the static potential, this construction yields renormalon-free short-distance coefficients with substantially reduced renormalization-scale sensitivity. The resulting reorganization of the perturbative expansion restores stable power counting and improves convergence across mesons, baryons, and tetraquarks. We have computed binding energies for heavy quarkonium, triply-heavy baryons, and all-heavy tetraquarks using Green's Function Monte Carlo methods applied to potential NRQCD with static potentials through next-to-next-to-leading order in the strong coupling (leading order in $1/m_Q$). The framework combines perturbatively calculated static potentials with MRS to stabilize the $\alpha_s$ expansion, then solves the resulting few-body Schr{\"o}dinger equations nonperturbatively via imaginary-time projection. Quark masses are extracted from spin-averaged $1S$ quarkonium masses (charmonium and bottomonium), then used without further tuning to predict baryon and tetraquark spectra.

At NLO and NNLO, scale variation induces shifts of order 100~MeV in bottomonium binding energies for fixed-order calculations ($\mu \in [\mu_p/2, 2\mu_p]$) but only $\sim 10\,\text{MeV}$ for MRS ($\lambda \in \{1/2, 1, 2\}$). At LO, the situation is reversed---this apparent enhancement reflects the inclusion of information about higher-order behavior already at LO in the MRS scheme, unlike in the fixed-order scheme where the LO potential is pure Coulomb and scale variation significantly underestimates perturbative truncation effects. 
For bottomonium, order-by-order shifts decrease from $\sim$20~MeV (LO$\to$NLO) to $\sim$2~MeV (NLO$\to$NNLO) in the MRS scheme, indicating that the short-distance series converges more rapidly when the leading renormalon is isolated into an explicit power correction.

Comparison with lattice QCD demonstrates both the successes and limitations of our static-potential treatment. For triply-heavy baryons, fixed-order NNLO undershoots lattice-QCD masses by 80--110~MeV across all four systems ($\Omega_{ccc}$, $\Omega_{ccb}$, $\Omega_{cbb}$, $\Omega_{bbb}$), while MRS NNLO undershoots by 125--175~MeV. The advantage of MRS is the reduced scale variation ($\sim 10$~MeV versus $\sim 800$~MeV) and smaller order-by-order shifts (below $\sim$20~MeV versus $\sim$20--55~MeV), which make the residual discrepancy unambiguous.
The fractional discrepancies decrease from 3.8\% for $\Omega_{ccc}$ to 1.0\% for $\Omega_{bbb}$, consistent with the interpretation that residual differences arise from neglected relativistic corrections (spin-orbit, spin-spin, Darwin terms), not from perturbative instability or incomplete potential matching. A direct test -- incorporating these corrections and verifying the discrepancy is absorbed -- is left for future work.

Our GFMC statistical uncertainties (1--10~MeV) are small compared to systematic uncertainties from missing $\mathcal{O}(1/m_Q)$ corrections. For bottom systems, these are estimated at $\sim \Lambda_\text{QCD}^2/m_b \sim 15$--20~MeV; for charm systems, the 125--175~MeV residual discrepancies with lattice QCD indicate that $1/m_Q$ corrections are the dominant source of uncertainty. Three immediate extensions would improve the accuracy and precision of our predictions. First, incorporating spin-dependent potentials at $\mathcal{O}(1/m_Q)$ and $\mathcal{O}(1/m_Q^2)$ should reduce the 125--175~MeV discrepancy with lattice QCD for charm and bottom systems. The structure of these operators and their associated potentials is well established within the pNRQCD framework~\cite{Brambilla:2004jw,Brambilla:2009bi,Pineda:2011dg}. While the potential formalism is known, the practical challenge lies in constructing trial wavefunctions that preserve GFMC efficiency while consistently encoding spin degrees of freedom. Second, genuine four-quark potentials arise at NNLO in the nonrelativistic expansion~\cite{Brambilla:2004jw,Brambilla:1999qa,Pineda:2000gza}, but remain incompletely constrained; their quantitative impact on tetraquark binding therefore remains uncertain. Third, extending the framework to excited-state spectroscopy requires improved projection strategies capable of isolating higher eigenstates within GFMC imaginary-time evolution, where signal degradation and state mixing present well-known difficulties.

We have presented binding energy predictions for a range of top-containing
hadrons -- toponium, $t\bar{c}$ and $t\bar{b}$ mesons, and tetraquarks such as 
$bb\bar{t}\bar{t}$, $bc\bar{t}\bar{t}$, and $cc\bar{t}\bar{t}$. Although these 
states cannot exist as asymptotic hadrons due to the top quark's rapid weak 
decay, they may produce observable signatures at the LHC or future colliders 
through near-threshold cross-section enhancements, modified angular correlations, 
or spin-dependent effects in $t\bar{t}$ production~\cite{Fuks:2021xje,Fuks:2025toponiumSimulation,Garzelli:2024uhe}.  The tetraquark sector 
provides an additional test of renormalon subtraction: binding thresholds for unequal-mass systems (Fig.~\ref{fig:unequalmass_binding_mrs}) shift from 
$m_Q/m_{Q'} \sim 0.145$ (LO) and $\sim 0.125$ (NLO) in fixed-order to $\sim 0.11$ (LO) and $\sim 0.10$ (NLO) in MRS.
At fixed-order NLO, the bound tetraquark states 
exhibit binding energies of 
500--600~MeV
below their respective two-quarkonium 
thresholds (Table~\ref{tab:tetra}).
Their color structure is consistent with a
diquark-antidiquark molecule picture for all-heavy tetraquarks.
Equal-mass $bb\bar{b}\bar{b}$ and $cc\bar{c}\bar{c}$ systems, and the insufficiently unequal-mass $cc\bar{b}\bar{b}$ where $m_c/m_b \approx 0.32$ exceeds the critical ratio -- show 
no evidence for bound states within our VMC/GFMC calculations.

The success of MRS in stabilizing heavy-hadron calculations extends renormalon subtraction to multi-body precision QCD. The consistent factor-of-10 improvement in scale dependence and the computational efficiency of coordinate-space QMC make pNRQCD-based predictions a useful complement to lattice QCD for heavy-quark systems. Several extensions are needed for full quantitative control: $1/m_Q$ relativistic corrections, MRS for non-singlet colored channels, and MRS resummation of three-body potentials. With these additions, the MRS-pNRQCD-QMC framework should provide a controlled perturbative description across the heavy-quark spectrum.

\acknowledgments
We thank Eric Braaten, Nora Brambilla, Tom Magorsch, and Antonio Vairo for helpful discussions and insightful comments. 
This manuscript has been authored by Fermi Forward Discovery Group, LLC under Contract No.\ 89243024CSC000002 with the U.S. Department of Energy, Office of Science, Office of High Energy Physics.
The work of BA is supported by the DOE grant No.\ DE-SC0011784 and NSF grant Nos.\ OAC-2103889, OAC-2411215, and OAC-2417682.
This work was performed in part at the Aspen Center for Physics, with support for BA by a grant from the Simons Foundation (1161654,Troyer).

\section*{Data availability}
The data that support the findings of this article are not publicly available.
The data are available from the authors upon reasonable request.

\appendix* % * removes the label "A" for a single Appendix
\section{Minimal renormalon subtraction}
\label{app:mrs}

Perturbative QCD calculations at fixed order suffer from factorial growth of coefficients in the $\alpha_s$ expansion, reflecting renormalon singularities in the Borel plane. This factorial behavior induces $\mathcal{O}(\Lambda_\text{QCD}/Q)$ ambiguities and large renormalization-scale dependence that worsen with increasing order. For the heavy-quark static potential, the leading renormalon manifests as $\sim 100$~MeV scale variation at NNLO. This is comparable to binding energies and undesirable for precision predictions.

Minimal renormalon subtraction~\cite{Komijani:2017vep,Brambilla:2017hcq,Kronfeld:2023jab} reorganizes the perturbative series to isolate and resum the renormalon tail, leaving behind short-distance coefficients with improved convergence. We summarize the formalism following Ref.~\cite{Kronfeld:2023jab}.

\subsection{MRS formalism}

In pNRQCD, static potentials are matching coefficient taking the form
\begin{equation}
    \mathcal{V}_R(r) = V_R(r) + \Lambda_R, \;\; V_R(r) = -\frac{C_R}{r} \sum_{l=0} v_l(\mu r) \alpha_s(\mu)^{l+1},
    \label{eq:R}
\end{equation}
where $R$ labels the irrep of the potential, $\Lambda_R$ is of order the QCD scale, and $v_l$ are perturbative coefficients.
The beta function coefficients satisfy
\begin{equation}
    \beta(\alpha_s) = -\alpha_s \sum_{k=0}^\infty \beta_k \alpha_s^{k+1}, \quad b \equiv \frac{\beta_1}{2\beta_0^2}.
\end{equation}
The presence of the constant $\Lambda$ of energy dimension~1 is related to the factorial growth of the $v_l$, and, in practice, the growth is clear from $v_l$, $l=0,\ldots,3$~\cite{Kronfeld:2023jab}.

Differentiating Eq.~(\ref{eq:R}) with respect to $r$ gives the static force
\begin{equation}
    \mathcal{F}(r) = \frac{C_R}{r^2} \sum_{k=0} f_k(\mu r) \alpha_s(\mu)^{k+1},
\end{equation}
with coefficients
\begin{equation}
    f_k = v_k - 2\sum_{l=0}^{k-1} (l+1)\beta_{k-1-l} v_l.
    \label{eq:fr-coeff}
\end{equation}
The derivative removes the power term and the leading factorial growth: $f_k$ grows more slowly than $v_k$ -- in fact, for the first four coefficients, no growth is observed whatsoever~\cite{Kronfeld:2023jab}.
Inverting Eq.~\eqref{eq:fr-coeff} yields
\begin{equation}
    v_l = f_l + \sum_{k=0}^{l-1} \frac{(k+1)\Gamma(l+1+b)}{\Gamma(k+2+b)} \left(2\beta_0\right)^{l-k} f_k.
    \label{eq:rrenorm}
\end{equation}
For large $l$ in principle, and any $l$ in practice, the second term dominates.

For $l \geq L$ (where $L$ is the highest computed order), a good approximation is thus $v_l \approx V_l$ with
\begin{align}
    V_l &= V_0 \left(2\beta_0\right)^l \frac{\Gamma(l+1+b)}{\Gamma(1+b)}, \\
    V_0 &= \sum_{k=0}^{L-1} (k+1) \frac{\Gamma(1+b)}{\Gamma(k+2+b)} \left(2\beta_0\right)^{-k} f_k .
    \label{eq:V0}
\end{align}
Separate the series into short-distance and renormalon parts:
\begin{equation}
    V(r) = V_\text{RS}(r) + V_\text{B}(r),
\end{equation}
where RS stand for ``renormalon subtraction'' and B for the Borel part.
We can express them as
\begin{align}
    V_\text{RS}(r) &= -\frac{C_R}{r} \sum_{l=0}^{L-1} (v_l - V_l)(\mu r) \alpha_s(\mu)^{l+1}, \\
    V_\text{B}(r)  &= -\frac{C_R}{r}\sum_{l=0}^\infty V_l \alpha_s(\mu)^{l+1}.
\end{align}
Note that $V_0$ is a function of $\mu r$, inherited from $f_k(\mu r)$, though for clarity we do not make this dependence explicit.
$V_\text{RS}(r; \mu r)$ and $V_\text{B}(r; \mu r)$ both depend on $\mu$, with the $\mu$ dependence canceling in the sum (in principle at all orders but in practice already at NNLO and N3LO)~\cite{Kronfeld:2023jab}.

The sum $V_\text{B}(r)$ can be interpreted via a Borel transformation:
\begin{equation}
    V_\text{B}(r) = -\frac{C_R}{r} \frac{V_0}{2\beta_0} \mathcal{J}(b, 1/(2\beta_0\alpha_s(\mu))),
    \label{eq:RB}
\end{equation}
with
\begin{equation}
    \mathcal{J}(c,y) = e^{-y} \Gamma(-c) \gamma^\star(-c,-y),
\end{equation}
where 
\begin{equation}
    \gamma^\star(a,x) = \frac{1}{\Gamma(a)} \int_0^1 dt\, t^{a-1} e^{-xt}.
\end{equation}
$\gamma^\star$ is known as the limiting function of the incomplete gamma function~\cite{AbramowitzStegun:1972}; it has a convergent expansion
\begin{equation}
    \gamma^\star(a,-y) = \frac{1}{\Gamma(a)} \sum_{n=0}^\infty \frac{y^n}{n!(n+a)},
\end{equation}
which converges rapidly even for $a = -b < 0$.

The final MRS prescription is
\begin{equation}
    \mathcal{V}(r) = V_\text{RS}(r) + V_\text{B}(r) + \Lambda_R,
    \label{eq:MRS_final}
\end{equation}
where the power correction coefficient $C_p$ would in principle be determined by 
fitting to (experimental or lattice) data.
In practice, when quark masses are extracted from experimental quarkonium masses as in Sec.~\ref{sec:qcd_mesons}, the power correction is implicitly absorbed into the extracted $m_Q$ and need not be specified separately.

\subsection{Implementation for pNRQCD potentials}

For the static potential $V(r,\mu)$, the MRS procedure described above is applied at each 
quark-antiquark separation $r$ with the identification $Q = 1/r$. The short-distance 
coefficients $v_l$ correspond to the perturbative expansion of $V(r,\mu)$ in powers of 
$\alpha_s(\mu)$, and the renormalon sum $R_\text{B}^{(p)}$ captures the leading infrared 
sensitivity with $p=1$.

At large separations $r \gtrsim 1$~fm, the natural scale $\mu \sim 1/r$ drops below 
$\Lambda_\text{QCD}$ and perturbative running of $\alpha_s(\mu)$ breaks down. Since MRS 
relies on the perturbative beta function to resum the renormalon tail, the procedure 
requires regularization in this regime. We introduce a cutoff scale $\mu_\text{cut}$ 
below which the coupling is regulated, and compare three implementation strategies:

\begin{itemize}
\item \textbf{Hard cutoff:} For $r > 1/\mu_\text{cut}$, revert to the 
fixed-order calculation with $\alpha_s$ frozen at $\alpha_s(\mu_\text{cut})$.

\item \textbf{Frozen $\alpha_s$:} Apply MRS at all $r$, but freeze 
$\alpha_s$ at $\alpha_s(\mu_\text{cut})$ whenever $\mu < \mu_\text{cut}$.

\item \textbf{$\mu'$:} Define a modified scale 
$\mu' = \sqrt{\mu_\text{cut}^2 + (1/r)^2}$ and use the running coupling $\alpha_s(\mu')$ 
everywhere. This interpolates smoothly between $\mu' \approx 1/r$ at short distances 
and $\mu' \approx \mu_\text{cut}$ at large distances.
\end{itemize}

\begin{figure} % [tb]
\centering
\includegraphics[width=0.98\linewidth]{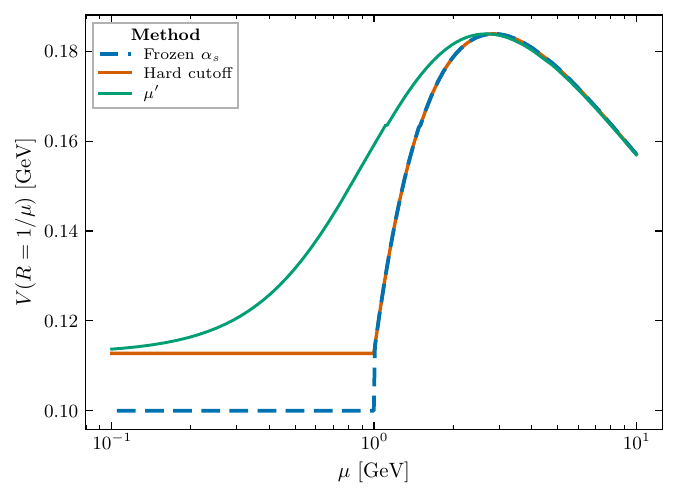}
\caption{Static potential $V(r)$ at NNLO for the three MRS regularization prescriptions with $\mu_\text{cut}=1$~GeV. 
The $\mu'$ prescription yields a smoothly varying potential with continuous derivatives across all $r$, 
up to the small discontinuity common to all schemes at $\mu = m_c$, where the number of active light flavors changes. 
In contrast, the hard-cutoff and frozen-$\alpha_s$ prescriptions display visible kinks near $r \sim 1/\mu_\text{cut}$.}
\label{fig:Comp_all_methods}
\end{figure}

\begin{figure} % [tb]
\centering
\includegraphics[width=0.98\linewidth]{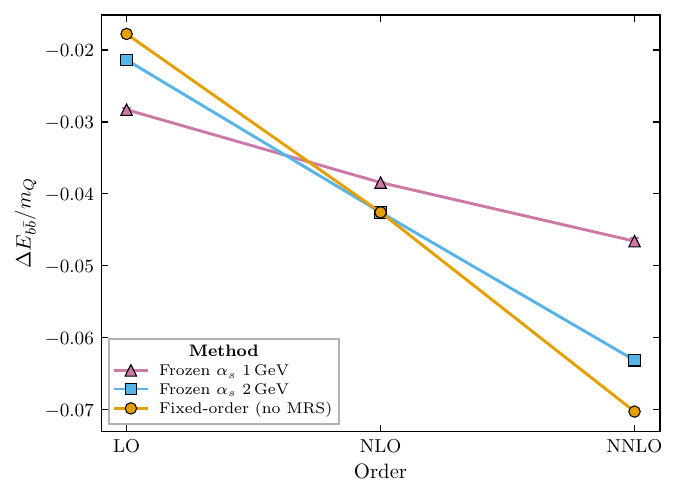}
\caption{Bottomonium binding energy versus renormalization scale factor comparing 
fixed-order (FO) results with hard cutoff at $\mu_\text{cut} = 1$~GeV and 
2~GeV. With $\mu_\text{cut} = 2$~GeV, typical quark separations satisfy $\mu > \mu_\text{cut}$ 
and MRS is rarely invoked, so results collapse to fixed-order. With $\mu_\text{cut} = 1$~GeV, 
MRS activates for typical heavy-quark separations and substantially reduces scale variation.}
\label{fig:cut_comparison}
\end{figure}

\begin{figure} % [!t]
\centering
\includegraphics[width=0.98\linewidth]{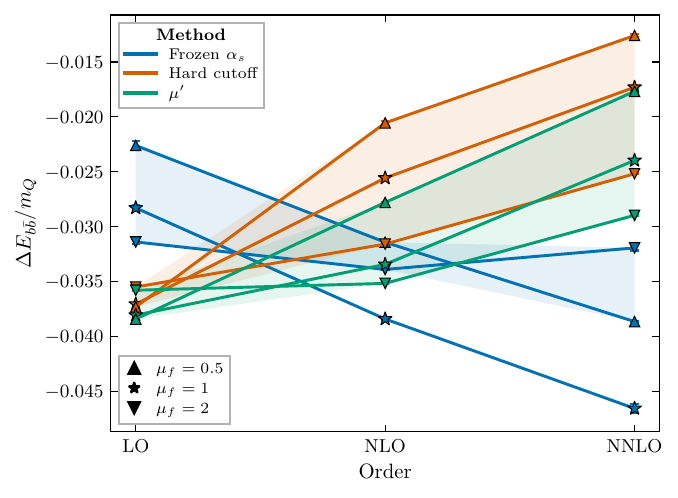}
\caption{Bottomonium binding energy comparing all three MRS implementation methods 
with $\mu_\text{cut} = 1$~GeV at LO, NLO, and NNLO. Lighter shades correspond to 
$\mu = \mu_p/2$; darker shades to $\mu = 2\mu_p$. $\mu'$ exhibits 
the smallest scale dependence, with residual variation of 10--20~MeV at NNLO compared 
to 100--150~MeV for fixed-order.}
\label{fig:1GeV_comparison}
\end{figure}

Figure~\ref{fig:Comp_all_methods} compares the resulting potentials. Hard cutoff and frozen $\alpha_s$ 
produce kinks in $V(r)$ near $r \sim 1/\mu_\text{cut}$ where the regularization activates, 
while $\mu'$ yields a smooth potential with continuous derivatives at all separations.

The choice of $\mu_\text{cut}$ affects how much of the integration region benefits from 
MRS. Figure~\ref{fig:cut_comparison} shows that with $\mu_\text{cut} = 2$~GeV, most 
Monte Carlo samples on determination of the bottomium meson probe $r < 0.5$~fm where $\mu > 2$~GeV, so MRS is rarely invoked 
and results collapse to fixed-order. With $\mu_\text{cut} = 1$~GeV, MRS activates over 
the physically relevant range of separations and reduces scale variation significantly.
We therefore adopt $\mu_\text{cut} = 1$~GeV for all calculations. The non-monotonic $\mu$-dependence visible in Fig.~\ref{fig:cut_comparison} reflects that the standard scale $\mu \sim \mu_p$ lies near a stationary point of the binding energy in $\mu$, where small $\mu$ variations cancel at leading order.

Figure~\ref{fig:1GeV_comparison} compares all three methods at $\mu_\text{cut} = 1$~GeV 
across perturbative orders. Fixed-order exhibits 100--150~MeV variation under 
$\mu \in [\mu_p/2, 2\mu_p]$ at NLO and NNLO. Hard cutoff retains 50--80~MeV variation 
with visible discontinuities. Frozen $\alpha_s$ improves to 30--50~MeV but retains residual 
$\mu$ sensitivity. $\mu'$ achieves 10--20~MeV variation -- an order of magnitude 
improvement over fixed-order.% -- with no derivative discontinuities.

We adopt $\mu'$ for all results presented in Secs.~\ref{sec:qcd_mesons}, \ref{sec:qcd_baryons}, and~\ref{sec:qcd_tetras}.
The smooth interpolation preserves the continuity of $\alpha_s(\mu')$, ensuring that both the potential and its gradients remain free of artificial kinks. 
This eliminates spurious force discontinuities in the GFMC evolution. 
The remaining ${\sim}10$~MeV scale variation probes the residual $\mu$-dependence at NNLO; it serves as a diagnostic of scheme stability rather than a quantitative truncation-error estimate.

\bibliography{pnrqcd, references_filtered}

\end{document}